\documentclass[12pt]{article}
\usepackage{amssymb,amsmath,amsthm,amsfonts,amscd}
\usepackage{graphicx}
\newcommand\be{\begin{equation}}
\newcommand\ee{\end{equation}}

\newcommand{\fatalpha}{{\bf \alpha \kern -0.44em \alpha}}
\newcommand{\fatsigma}{{\bf \sigma \kern -0.54em \sigma}}
\newcommand{\tpchi}{{\bf \chi \kern -0.35em \chi}}
\newcommand{\llambda}{{\bf \lambda \kern -0.45em \lambda}}

\bibliography{plain}
 \title{\bf  Entanglement of Multi-qudit States Constructed
by Linearly Independent Coherent States: Balanced Case}
 \vspace{20mm}
\author{G. Najarbashi \thanks{E-mail: Najarbashi@uma.ac.ir} ,
  S. Mirzaei \thanks{E-mail: SMirzaei@uma.ac.ir}
\\
{\small Department of Physics, University of Mohaghegh Ardabili, P.O. Box 179, Ardabil, Iran.} \\
 \\
}\pagebreak
\begin{document}
\maketitle \vspace{0mm}

\maketitle \vspace{0mm}
\begin{abstract}
Multi-mode entangled coherent states are important resources for linear optics quantum computation and  teleportation.
Here we introduce the generalized balanced N-mode coherent states which recast in  the multi-qudit case. The necessary and sufficient condition for bi-separability of such balanced N-mode coherent states is found.
We particularly focus on pure and mixed multi-qubit and multi-qutrit like states and examine the degree of bipartite
as well as tripartite entanglement using the concurrence measure. Unlike the N-qubit case, it is shown that there are
 qutrit  states  violating monogamy inequality.
Using parity, displacement operator and beam splitters,  we will propose a scheme for generating balanced N-mode   entangled coherent states for  even number of terms in superposition.
\end{abstract}
\newpage
\section{Introduction}
Coherent states, originally introduced by Schr¨odinger in 1926 \cite{Schrodinger},  refer to a special kind of pure quantum-mechanical state of the light field corresponding to a single resonator mode which describe closest quantum state to a classical sinusoidal wave such as a continuous laser wave. However, for multi-mode fields, the prospects for nonclassical effects
become even richer as long as the field states are not merely product states of each of the modes which means that  the multi-mode is entangled state \cite{EPR}.
\par
Entangled coherent states have many applications in quantum optics and quantum information processing \cite{Cochrane,Oliveira,Kim,Barry,Milburn,Munro,wang5,Enk,wang3,Vogel1}. Communication via
entangled coherent quantum network is investigated in \cite{Allati1} where it is shown that the probability
of performing successful teleportation through this network depends on its size. The nonlinear Mach-Zehnder interferometer is presented as a device whereby a pair of coherent states can be transformed into an entangled superposition of coherent states for which the notion of entanglement is generalized to include nonorthogonal, but distinct, component states \cite{Barry,Cheong1,Enk1}.
In \cite{wang1}, it is proposed a scheme for generating multipartite entangled coherent states via entanglement swapping, with an example of a physical realization in ion traps and then  bipartite entanglement of these multipartite states is quantified by the concurrence.
The required conditions for the maximal entanglement in superposed bosonic coherent
states of the form
\be
|\psi\rangle=\mu|\alpha\rangle|\beta\rangle+\nu|\gamma\rangle|\delta\rangle,
\ee
have been studied in references  \cite{wang2,wang4}, and subsequently have been
generalized to the state
\be
 |\psi\rangle=\mu|\alpha\rangle|\beta\rangle+\lambda|\alpha\rangle|\delta\rangle+
 \rho|\gamma\rangle|\beta\rangle+\nu|\gamma\rangle|\delta\rangle,
\ee
 in Ref. \cite{najarbashi}.
   In Ref. \cite{DAOUD1} the generation of multipartite entangled $SU(k+1)$ coherent states using a quantum network involving a sequence of k beam splitters have been investigated. Based on Glauber coherent states, the even and odd
three-mode Schrodinger cat states and limitations to
sharing quantum correlations known as monogamy relations have been investigated in  Ref.   \cite{DAOUD2}.
 \par
In this paper we consider the generalized balanced  N-mode Glauber coherent states of the form
\be\label{mainstate}
|\Psi^{(d)}_{N}\rangle=\frac{1}{\sqrt{M^{(d)}_N}}\sum_{i=0}^{d-1}\mu_i|\underbrace{\alpha_i\rangle\cdots|\alpha_i\rangle}_{\mathrm{N \ modes}},
\ee
which is a general form of  the balanced two-mode entangled coherent state $|\psi\rangle_{bal}=\frac{1}{\sqrt{N}}(|\alpha\rangle|\alpha\rangle+|\beta\rangle|\beta\rangle)$. As two coherent states are in general nonorthogonal, they  span a two dimensional qubit like Hilbert space $\{ |0\rangle, |1\rangle\}$. Therefore, balanced two-mode coherent state $|\psi\rangle_{bal}$ can  be recast in two qubit form $\sum_{i,j=0,1}a_{ij}|ij\rangle$.
The same argument can be formulated for other two modes coherent states such as
\be
|\psi'\rangle_{bal}=\frac{1}{\sqrt{N'}}(|\alpha\rangle|\alpha\rangle+|\beta\rangle|\beta\rangle+|\gamma\rangle|\gamma\rangle),
\ee
if  the set $\{ |\alpha\rangle,|\beta\rangle, |\gamma\rangle \}$ are linearly independent, i.e. when  they span the three dimensional qutrit like Hilbert space  $\{ |0\rangle, |1\rangle\,|2\rangle \}$. Hence $|\psi'\rangle_{bal}$ can be recast in two qutrit form $|\psi'\rangle_{bal}=\sum^2_{i,j=0}a_{ij}|ij\rangle$.
One may proceed in the same manner and represent the generalized balanced  N-mode Glauber coherent states $|\Psi^{(d)}_{N}\rangle$
as the multipartite qudit  state
\be
|\Psi^{(d)}_{N}\rangle=\sum^{d-1}_{i_0i_1...i_{d-1}=0}a_{i_0i_1...i_{d-1}}|i_0i_1...i_{d-1}\rangle,
\ee
if the set $\{ |\alpha_i\rangle\}^{d-1}_{i=0}$ are linearly independent.  The entanglement of the state $|\Psi^{(d)}_{N}\rangle$  can be calculated, using some useful measure such as concurrence, in the above multipartite qudit form.
\par
The outline of this paper is as follows: In section 2 we begin with a rather simple case, i.e. multi-mode qubit like coherent states. Using concurrence measure we find the necessary and sufficient condition for maximally bipartite entanglement of these sates. Furthermore, mixed states and monogamy inequality is investigates in this section. Section 3 devoted to the pure and mixed multi-qutrit case. It is shown that there are qutrit like states in which the monogamy inequality is violated. The necessary and sufficient condition for separability of generalized balanced multi-mode  coherent states is found in section 4.
In section 5, we propose a scheme to produce the generalized balanced N-mode  coherent states with superposition of even terms.
Concluding remarks close this paper.
\par
\section{Multi-qubit case}
We begin our discussion by recapitulating some fundamental notions of coherent states
, as formulated in the traditional language of creation and annihilation operators $\hat{a}$ and $\hat{a}^\dag$ as
\be\label{displacement}
|\alpha\rangle=\hat{D}(\alpha)|0\rangle=\exp(\alpha \hat{a}^{\dag}-\alpha^{*}\hat{a})|0\rangle,
\ee
where $\hat{D}(\alpha)$ called displacement operator and $\alpha$ is an arbitrary complex number. Coherent state (also called Glauber state) refer to a special kind of pure quantum-mechanical state of the light field corresponding to a single resonator mode which describe closest quantum state to a classical sinusoidal wave such as a continuous laser wave.
Alternatively, the coherent states are eigenstates of annihilation operators, i.e. $\hat{a}|\alpha\rangle=\alpha|\alpha\rangle$. The displacement operator is unitary with  $\hat{D}^\dag(\alpha)=\hat{D}^{-1}(\alpha)=\hat{D}(-\alpha)$ and  multiplication rule
\begin{equation}
\hat{D}(\alpha)\hat{D}(\beta)=e^{i Im(\alpha \beta^*)}\hat{D}(\alpha+\beta).
\end{equation}
Expanding the exponential in displacement operator and using the relations $\hat{a}|n\rangle=\sqrt{n}|n-1\rangle$ and $\hat{a}^\dag|n\rangle=\sqrt{n+1}|n+1\rangle$ one can write the coherent states in terms of Fock states $|n\rangle$ as
\be
|\alpha\rangle=e^{\frac{-|\alpha|^{2}}{2}}\sum_{n=0}^\infty\frac{\alpha^{n}}{\sqrt{n!}}|n\rangle.
\ee
On the other hand, using, the  orthonormality of Fock states  ($\langle n|m\rangle=\delta_{nm}$),  the overlap of two coherent states reads
\be\label{overlap}
\langle\alpha|\beta\rangle=e^{-\frac{1}{2}(|\alpha|^{2}+|\beta|^{2}-2\alpha^{*}\beta)}.
\ee
\par
Let us now consider an N-mode state $|\Psi^{(2)}_{N}\rangle$ expressed in terms of  coherent states $|\alpha\rangle$ and $|\beta\rangle$ as follows
\be\label{4}
|\Psi^{(2)}_{N}\rangle=\frac{1}{\sqrt{M^{(2)}_{N}}}(|\alpha\rangle|\alpha\rangle\cdots|\alpha\rangle+\mu|\beta\rangle|\beta\rangle\cdots|\beta\rangle),
\ee
where $\mu$, $\alpha$ and $\beta$  are generally complex numbers and $M_{N}$ is a factor normalizing $|\Psi^{(2)}_{N}\rangle$, i.e.
\begin{equation}
M^{(2)}_N=1+|\mu|^{^2}+2Re(\mu p^N),
\end{equation}
in which $p=\langle\alpha|\beta\rangle$ is equal to Eq. (\ref{overlap}). Note that we  used the superscript $(2)$ for qubit-like states to distinguish it from that of qutrit like or other states in the next section. Two non-orthogonal coherent states $|\alpha\rangle$ and
 $|\beta\rangle$ are assumed to be linearly independent and span a two-dimensional subspace of the Hilbert space. The state $|\Psi^{(2)}_{N}\rangle$ is in general an entangled state. To see this, all that we have to do is to evaluate its concurrence \cite{Wootters1,Wootters2,Akhtarshenas}. Recall that a general form of  bipartite quantum state in the usual orthonormal basis $|e_{i}\rangle$ is
\be\label{state1}
|\psi\rangle=\sum_{i=1}^{d_{1}}\sum_{j=1}^{d_{2}}a_{ij}|e_{i}\otimes e_{j}\rangle,
\ee
where $d_{1}$ and $d_{2}$  are  dimensions of first and second part respectively. The norm of concurrence vector is defined as
\be
C=\sqrt{\sum_{a=1}^{d_{1}(d_{1}-1)/2}~~\sum_{b=1}^{d_{2}(d_{2}-1)/2}|C_{ab}|^{2}},
\ee
where $C_{ab}=\langle\psi|\widetilde{\psi}_{ab}\rangle$, $|\widetilde{\psi}_{ab}\rangle=(L_{a}\otimes L_{b})|\psi^{*}\rangle$, and $L_{a}$ and $L_{b}$ are the generators of $SO(d_{1})$ and $SO(d_{2})$ respectively. Note that $|\psi^{*}\rangle$ is complex conjugate of  $|\psi\rangle$.
The concurrence in terms of coefficients $a_{ij}$ is
\be\label{cc}
C = 2  \sqrt {\sum\limits_{i < j}^{d_1 } {\sum\limits_{k < l}^{d_2 } {\left| {a_{ik} a_{jl}  - a_{il} a_{jk} } \right|^2 } } }.
\ee
Moreover, when the state (\ref{state1}) is maximally entangled then $C$ takes its maximum value $\sqrt{2(d-1)/d}$ with $d=\min\{d_{1},d_{2}\}$.
Returning to the particular problem at hand, one can transform the  $|\Psi^{(2)}_{N}\rangle$  to a form analogous
to Eq. (\ref{state1}) by defining the orthonormal basis
\be\label{base}
|0\rangle=|\alpha\rangle,\quad \quad
|1\rangle=\frac{|\beta\rangle-p|\alpha\rangle}{N_{1}},
\ee
where $N_{1}=\sqrt{1-|p|^{2}}$. As the state $|\Psi^{(2)}_{N}\rangle$ is symmetric, i.e. its form is unchanged by permuting  each two modes, thus, without lose of generality, we can assume  the first $m$ modes as part one and the  remaining  $(N-m)$ modes as the second part which implies that it can be represented in a simple two-qubit like form
\be\label{sai}
\begin{array}{l}
|\Psi^{(2)}_{N}\rangle=\frac{1}{\sqrt{M^{(2)}_N}}[(1+\mu p^N)|\mathbf{00'}\rangle+\mu p^{m}(\sqrt{1-|p|^{2(N-m)}})|\mathbf{01'}\rangle+\mu p^{N-m}(\sqrt{1-|p|^{2m}})|\mathbf{10'}\rangle\\
~~~~+\mu\sqrt{(1-|p|^{2m})(1-|p|^{2(N-m)})}|\mathbf{11'}\rangle],
\end{array}
\ee
where
\be
\begin{array}{l}
|\mathbf{0}\rangle\equiv\underbrace{|0\rangle\cdots|0\rangle}_{m},\quad\quad|\mathbf{0'}\rangle\equiv\underbrace{|0\rangle\cdots|0\rangle}_{N-m},\\
|\textbf{1}\rangle\equiv\underbrace{|1\rangle\cdots|1\rangle}_{m},\quad\quad|\mathbf{1'}\rangle\equiv\underbrace{|1\rangle\cdots|1\rangle}_{N-m}.
\end{array}
\ee
As for general two-qubit pure state $|\psi\rangle=a|00\rangle+b|01\rangle+c|10\rangle+d|11\rangle$ the concurrence (\ref{cc}) reduces to $C=2|ad -bc|$, one can immediately deduce that  Eq. (\ref{4}) has concurrence
\be\label{conc1}
{C^{(2)}_{m,N-m}}=\frac{2|\mu||(1-|p|^{^{2m}})(1-|p|^{^{2(N-m)}})|^{^\frac{1}{2}}}{1+|\mu|^{^2}+2R|\mu|\cos(\phi+A)
},
\ee
where
\begin{equation}
A=N|\alpha||\beta|\sin(\theta_2-\theta_1),\quad R=\exp[\frac{-N}{2}(|\alpha|^{^2}+|\beta|^{^2}-2|\alpha||\beta|\cos(\theta_2-\theta_1))],
\end{equation}
in which we have defined  $\mu=|\mu|e^{i\phi}$, $\alpha=|\alpha|e^{i\theta_{1}}$ and $\beta=|\beta|e^{i\theta_{2}}$. We  used the superscript $(2)$ for concurrence of qubit-like states to distinguish it from that of qutrit like or other states in the next section.
Before proceeding with explicit examples, let us pause to make some preliminary remarks on  concurrence (\ref{conc1}).
The multi-mode (\ref{4}) is separable if and only if $\mu=0$ or $p=1$. The former is trivial and the latter implies  that the state $|\alpha\rangle$, is equal to $|\beta\rangle$, up to a phase factor, which is forbidden by our early assumption, i.e. linearly independent of $|\alpha\rangle$ and $|\beta\rangle$. On the other hand, maximal entangled states can be obtained by solving the equation $C=1$ or equivalently the equation $|\mu|^2+2|\mu|(R\cos(\phi+A)+R-1)+1=0$ must be solve for $|\mu|$. This in turn requires that $N=2m$. Since this equation has negative discriminant,  $\mu,\alpha$ and $\beta$ must be real numbers.
\par
Now, let us suppose that $\alpha$, $\beta$ and $\mu$ are real numbers and for simplicity we consider two-mode state
$|\Psi^{(2)}_{2}\rangle=\frac{1}{\sqrt{M^{(2)}_N}}(|\alpha\rangle|\alpha\rangle+\mu|\beta\rangle|\beta\rangle)$
with concurrence
\be\label{1}
C^{(2)}_{1, 1}=\frac{2|\mu(1-p^{2})|}{|1+\mu^{2}+2\mu p^{2}|},
\ee
which has  maximum (i.e. $C=1$) in $\mu=-1,$ with $0\leq p<1$ and $\mu=1$, with $ p=0$ (see figure 1).
\begin{figure}
\centerline{\includegraphics[width=8cm]{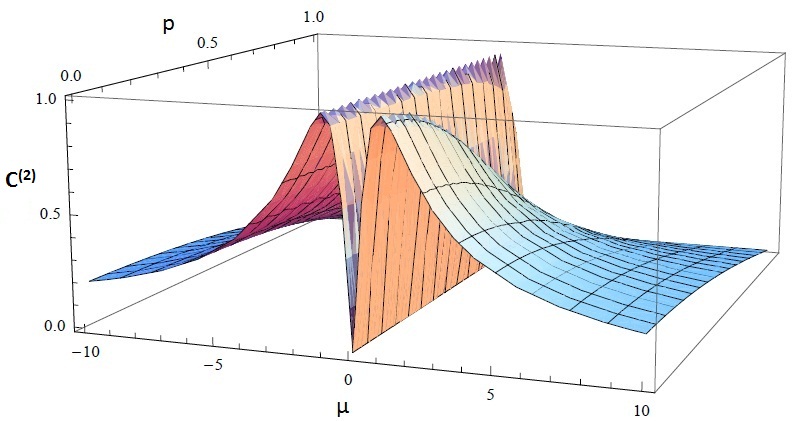}}
 \caption{\small Concurrence of state $|\Psi^{(2)}_{N}\rangle$  as a function of $\mu$ and $p$}\label{1}
\end{figure}
Figure 2 indicates the behaviour  of concurrence for different  values of $p=0.1, 0.3$ and $ 0.5$ which shows that for $\mu>0$ the concurrence significantly decreases by increasing $p$, while for  $\mu<0$ the changes are insignificant.
\begin{figure}
\centerline{\includegraphics[width=10cm]{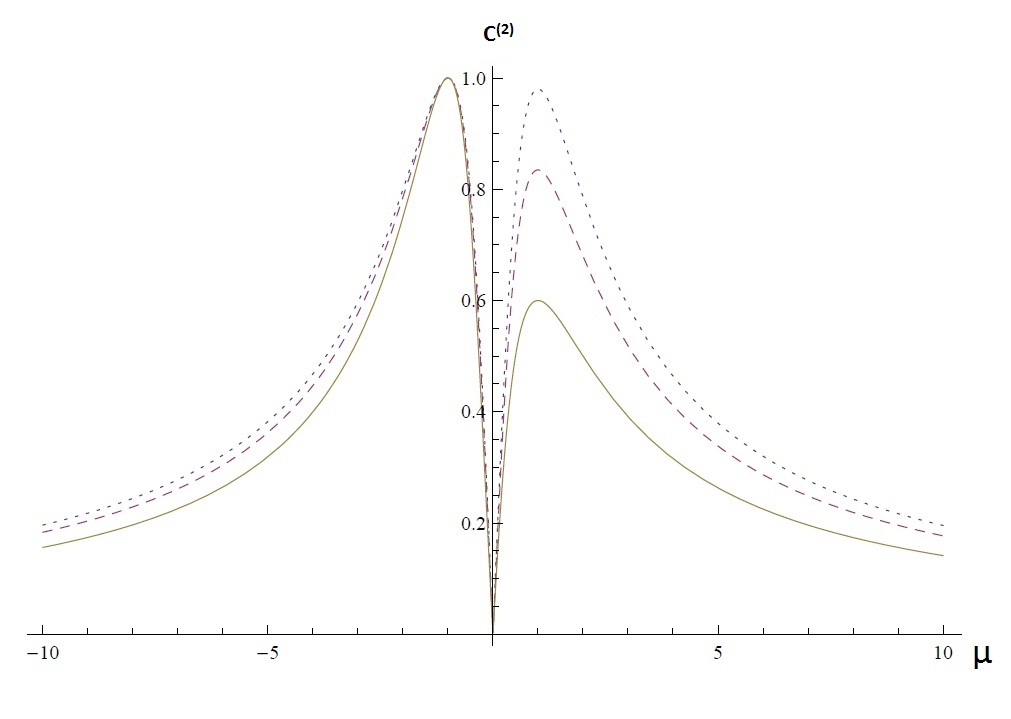}}
 \caption{\small Concurrence $C^{(2)}_{1, 1}$  as a function of $\mu$ for different values:  $p_{1}=0.1$ (dotted line), $p_{2}=0.3$ (dashed line) and  $p_{3}=0.5$ (full line).}
\end{figure}
A profile of concurrence is depicted in figure 3 as a function of $p$ in  range $[0,1]$ for different  values of positive $\mu$. As $\mu$ is increased, the concurrence grows to attain its maximum value $C=1$ when  $p$ tend to zero namely  $|\alpha\rangle$  and $|\beta\rangle$ are orthogonal ($\alpha \ \mathrm{or}\ \beta \rightarrow\infty$).
\begin{figure}
\centerline{\includegraphics[width=8cm]{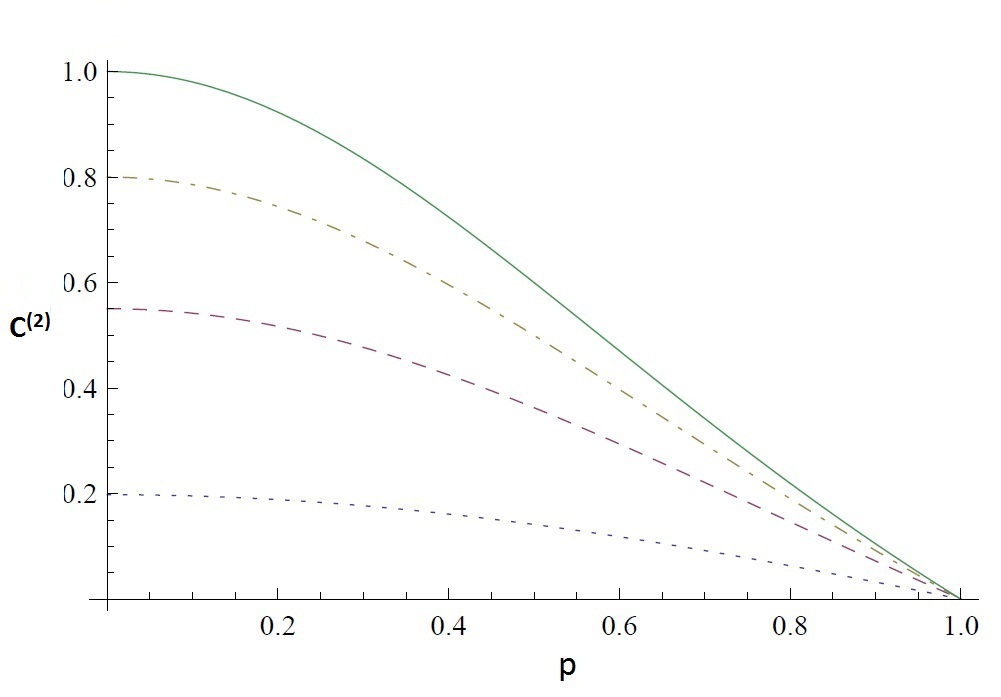}}
 \caption{\small Concurrence $C^{(2)}_{1, 1}$  as a function of $p$, for different values of  $\mu=0.1$ (dotted line), $\mu=0.3$ (dashed line),  $\mu=0.5$ (dot-dashed line) and  $\mu=1$ (full line).}
\end{figure}
It is instructive to compare the concurrence of the states $|\Psi^{(2)}_{N}\rangle$,  $|\Psi^{(3)}_{N}\rangle$, $|\Psi^{(4)}_{N}\rangle$ and $|\Psi^{(5)}_{N}\rangle$ with fixed $m=1$  but several different values $N=2,3,4$ and $5$ respectively (see figure 4). We found that the concurrence  is more sensitive for positive $\mu$ than for negative one.
\begin{figure}
\centerline{\includegraphics[width=11cm]{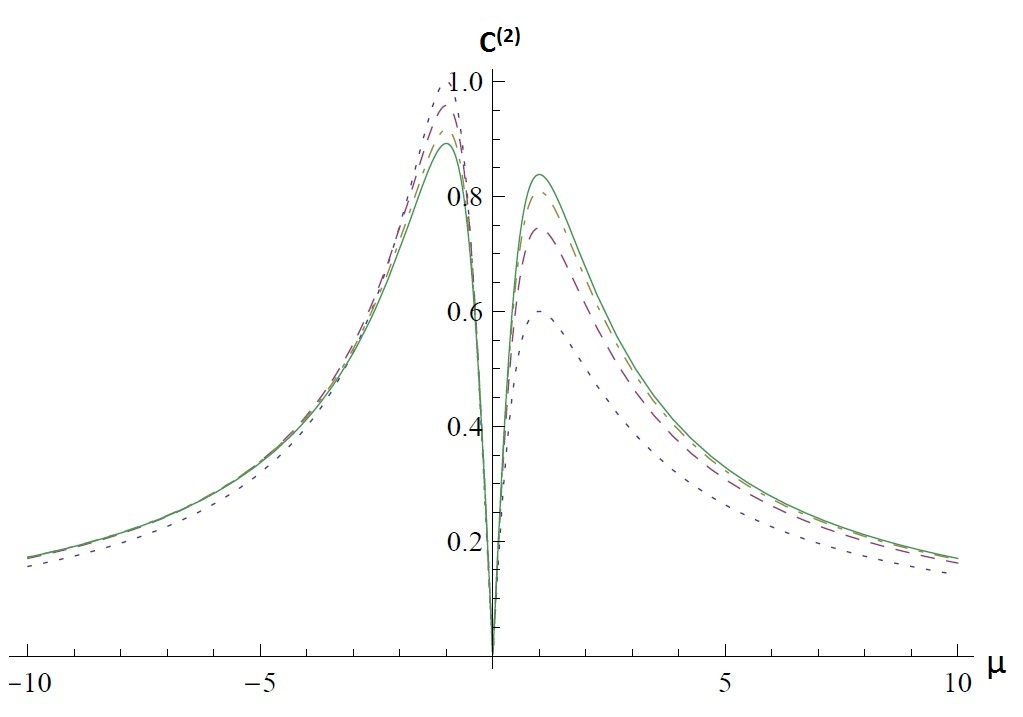}}
 \caption{\small Concurrences  $C^{(2)}_{1,1}$ (dotted line),  $C^{(2)}_{1,2}$ (dashed line),  $C^{(2)}_{1,3}$ (dot-dashed line), $C^{(2)}_{1,4}$ (full line) as a function of $\mu$ in the range $[-10,10]$.}\label{2}
\end{figure}
Similarly, one may fix $N$ and $p$, (e.g. ten modes  with $p=0.8$ in or example) and explore the behaviour of concurrence as a function of $\mu$. There is significant enhancement of the concurrence by the increasing $m$($\leq\frac{N}{2}$) where the maximum is achieved for balanced partition, i.e. $m=\frac{N}{2}$  (see figure 5).
\begin{figure}
\centerline{\includegraphics[width=10cm]{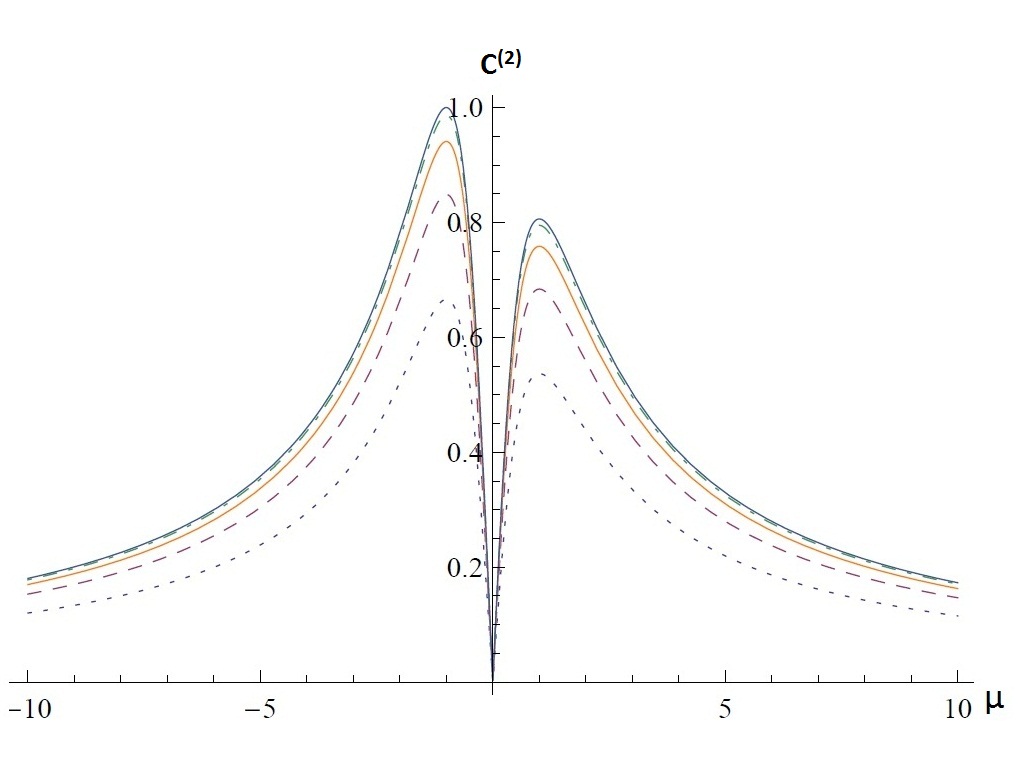}}
  \caption {\small Concurrences  $C^{(2)}_{1,9}$ (dotted line),  $C^{(2)}_{2,8}$ (dashed line),  $C^{(2)}_{3,7}$ (orange line), $C^{(2)}_{4,6}$ (dot-dashed line), $C^{(2)}_{5,5}$ (full line)  as a function of $\mu$ in the range $[-10,10]$ for fixed number of mode $N=10$ and given $p=0.8$.}\label{3}
\end{figure}
\subsection{Mixed states and monogamy inequality for multi-qubit case}
\begin{figure}
\centerline{\includegraphics[width=10cm]{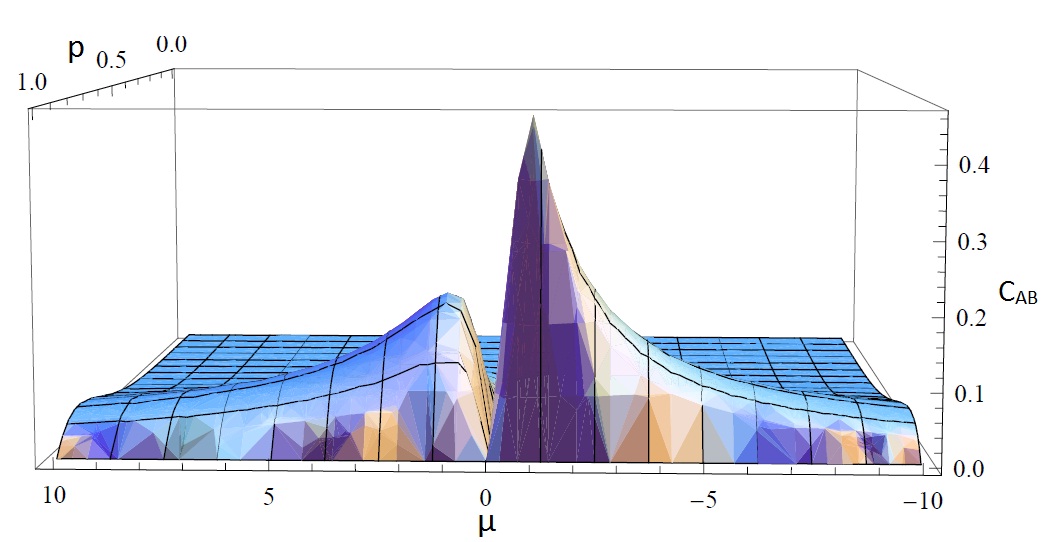}}
  \caption{\small Concurrences  $C_{AB}$ as a function of $\mu, p$ for fixed number of mode $N=10$, $m_{1}=2$ and $m_{2}=3$.}\label{cab}
\end{figure}

\begin{figure}
\centerline{\includegraphics[width=16cm]{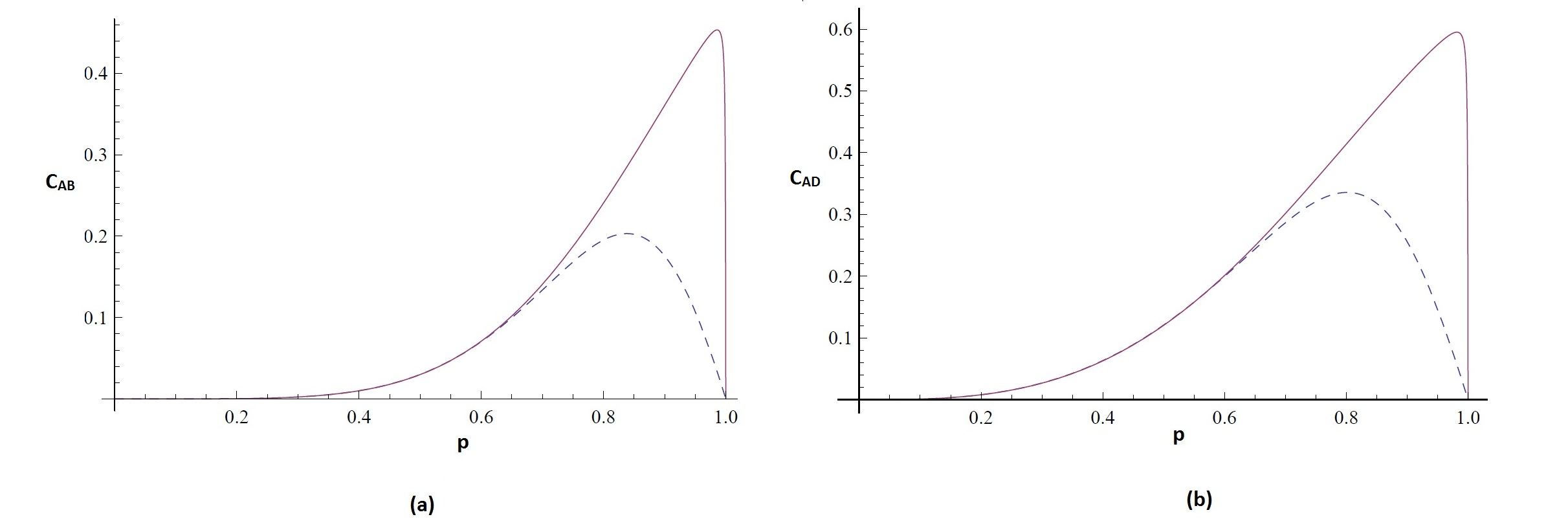}}
  \caption{\small (a) Concurrences  $C_{AB}$ as a function of $p$ for $\mu=-1$(full line) and $\mu=1$ (dashed line) ( for fixed number of modes $N=10$, $m_{1}=2$ and $m_{2}=3$). (b) The same profile for concurrence $C_{AD}$. }\label{profile1}
\end{figure}

\begin{figure}
\centerline{\includegraphics[width=10cm]{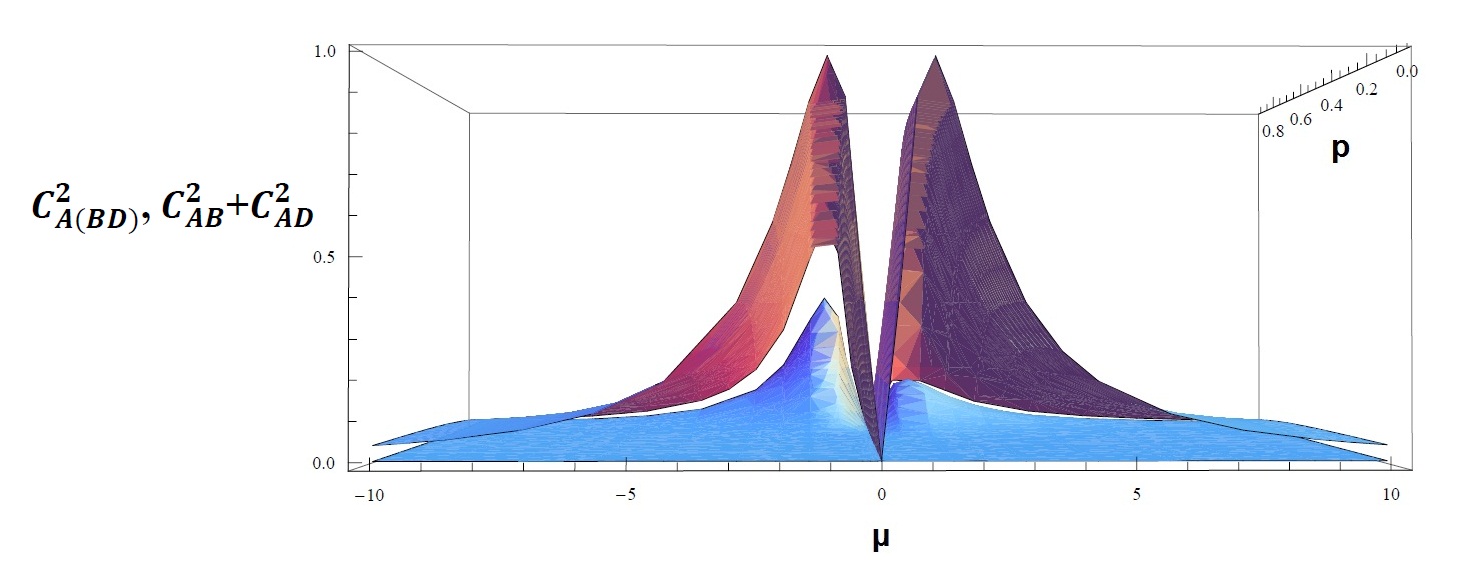}}
  \caption{\small  $C^2_{A(BD)}$ (upper surface) and $C_{AB}^2+C_{AD}^2$ (lower surface) as  functions of $\mu$ and $ p$ for fixed number of modes $N=10$, $m_{1}=2$ and $m_{2}=3$.}\label{mono1}
\end{figure}

\begin{figure}
\centerline{\includegraphics[width=10cm]{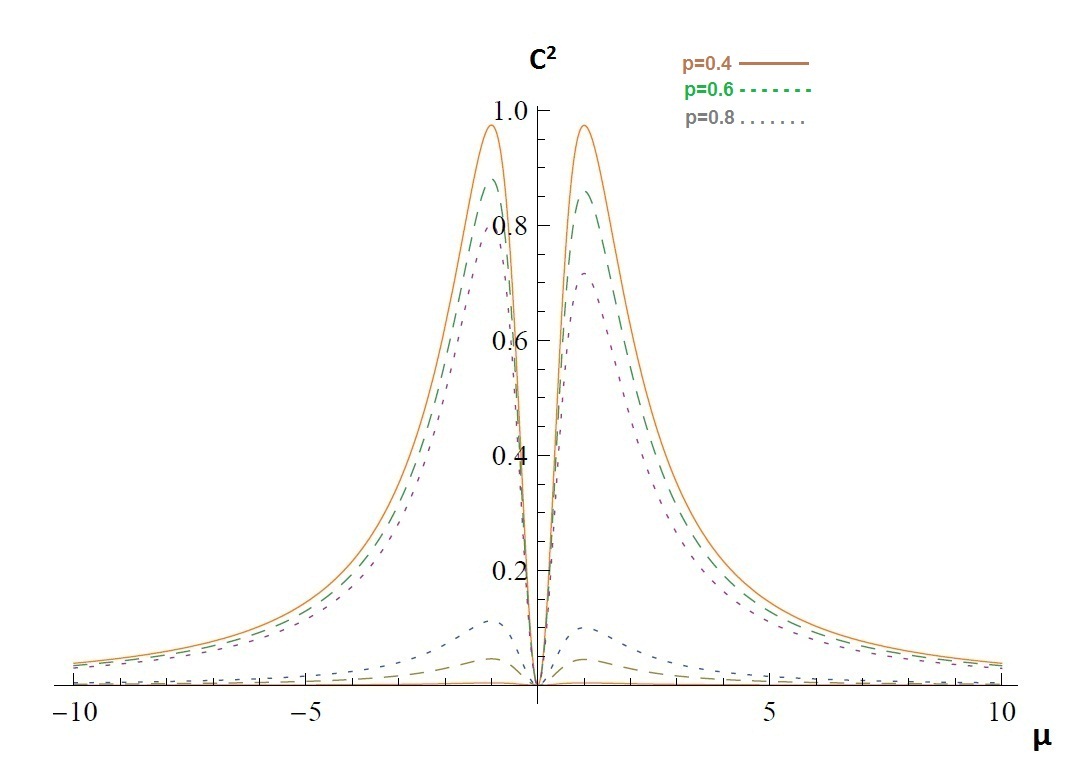}}
  \caption{\small Concurrences as a function of $\mu$ for fixed number of mode $N=10$, $m_{1}=2$, $m_{2}=3$ and $p=0.8$. $C^{2}_{AB}+C^{2}_{AD}$(dashed line) and $C^{2}_{A(BD)}$ (full line) }\label{profile2}
\end{figure}

Returning to the prototype multi-mode state (\ref{4}) discussed earlier, let us focus on entanglement of  mixed states that arise from (\ref{4}) by partially tracing out some subsystem. To this end, let us consider the state (\ref{4}) as three partitions $A,B$ and $D$ including $m_1,m_2$ and $m_3=N-m_1-m_2$ modes respectively, viz
\be
\begin{array}{l}
|\Psi^{(2)}_{N}\rangle_{ABD}=\frac{1}{\sqrt{M^{(2)}_N}}[(1+\mu p^N)|\textbf{00}'\textbf{0}''\rangle+\mu N''_{1}p^{m_{1}+m_{2}}|\textbf{00}'\textbf{1}''\rangle+\mu N'_{1}p^{N-m_{2}}|\textbf{01}'\textbf{0}''\rangle+\mu N_{1}p^{N-m_{1}}|\textbf{10}'\textbf{0}''\rangle\\
~~~~~+\mu N'_{1} N''_{1}p^{m_{1}}|\textbf{01}'\textbf{1}''\rangle+\mu N_{1} N''_{1}p^{m_{2}}|\textbf{10}'\textbf{1}''\rangle+\mu N_{1} N'_{1}p^{N-m_{1}-m_{2}}|\textbf{11}'\textbf{0}''\rangle+\mu N_{1}N'_{1} N''_{1}|\textbf{11}'\textbf{1}''\rangle],
\end{array}
\ee
where $N_{1}=\sqrt{1-p^{2m_{1}}}$, $N'_{1}=\sqrt{1-p^{2m_{2}}}$, $N''_{1}=\sqrt{1-p^{2 m_{3}}}$ and
\be
\begin{array}{l}
|\mathbf{0}\rangle\equiv\underbrace{|0\rangle\cdots|0\rangle}_{m_{1}},\quad\quad|\mathbf{0}'\rangle\equiv\underbrace{|0\rangle\cdots|0\rangle}_{m_{2}},\quad\quad|\mathbf{0}''\rangle\equiv\underbrace{|0\rangle\cdots|0\rangle}_{m_{3}},\\
|\mathbf{1}\rangle\equiv\underbrace{|1\rangle\cdots|1\rangle}_{m_{1}},\quad\quad|\mathbf{1}'\rangle\equiv\underbrace{|1\rangle\cdots|1\rangle}_{m_{2}},\quad\quad|\mathbf{1}''\rangle\equiv\underbrace{|1\rangle\cdots|1\rangle}_{m_{3}}.
\end{array}
\ee
 With this decomposition at hand, we can obtain the reduced density matrices $\rho_{AB}=Tr_{D}(|\Psi^{(2)}_{N}\rangle_{_{ABD}}\langle\Psi^{(2)}_{N}|)$ and $\rho_{AD}=Tr_{B}(|\Psi^{(2)}_{N}\rangle_{_{ABD}}\langle\Psi^{(2)}_{N}|)$ by tracing out subsystems $D$ and $B$ respectively. In general the density matrices  $\rho_{AB}$ and  $\rho_{AD}$ are mixed.
For any two-qubit mixed state, concurrence is defined as $C=\max\{0,\lambda_{1}-\lambda_{2}-\lambda_{3}-\lambda_{4}\}$ (concentrating on two-qubit  basis we suppress the superscript $(2)$ throughout) where the $\lambda_{i}$'s are the non-negative eigenvalues, in decreasing order, of the Hermitian matrix $R=\sqrt{\sqrt{\rho}\tilde{\rho}\sqrt{\rho}}$,
with $\tilde{\rho}=(\sigma_{y}\otimes\sigma_{y})\rho^{*}(\sigma_{y}\otimes\sigma_{y})$
in which $\rho^{*}$ is the complex conjugate of $\rho$ when it is expressed in a standard basis and $\sigma_{y}$ represents the usual second Pauli matrix in a local basis $\{|0\rangle, |1\rangle\}$ \cite{Wootters1}.
We can fix  $N$, $m_1$ and $m_2$ (e.g. $N=10$, $m_{1}=2$ and $m_{2}=3$) and explore monogamy inequality  \cite{coffman,kim,sanders1}
\be
\label{monogamy}
C_{A(BD)}^{2}\geq C_{AB}^{2}+C_{AD}^{2},
\ee
where $C_{AB}$ and $C_{AD}$ are the concurrences of the reduced density matrices of $\rho_{AB}$ and $\rho_{AD}$ respectively and
$C_{A(BD)}$ is the concurrence of pure state $|\Psi^{(2)}_{N}\rangle_{ABD}$ with respect to the partitions  $A$ and $BD$. The reduced density matrix $\rho_{AB}$ has the form
\be
 \rho_{AB}=\frac{1}{M^{(2)}_N}
 \left(
   \begin{array}{cccc}
     \rho_{11} & \rho_{12} & \rho_{13} & \rho_{14} \\
     \rho_{12} & \rho_{22} & \rho_{23} & \rho_{24} \\
     \rho_{13} & \rho_{23} & \rho_{33} & \rho_{34}  \\
     \rho_{14} & \rho_{24} & \rho_{34} & \rho_{44} \\
   \end{array}
 \right),
 \ee
with
\be
\begin{array}{l}
    \rho_{11}= 1+2\mu p^N+\mu^2 p^{2(m_1+m_2)},\ \quad  \rho_{12}=\mu N'_{1}(p^{N-m_2}+\mu p^{2m_1+m_2}), \\
    \rho_{13}=\mu N_1(p^{N-m_1}+\mu p^{m_1+2m_2}),\ \quad \rho_{14}=\mu N_1 N'_1(p^{m_3}+\mu p^{m_1+m_2}),\\
    \rho_{22}=\mu^{2}N'^{2}_{1}p^{2m_{1}},\ \quad \rho_{23}=\mu^2 N_1 N'_1 p^{m_1+m_2},\ \quad \rho_{24}=\mu^2 N_1 N'^{2}_1 p^{m_1}, \\
    \rho_{33}=\mu^{2}N^{2}_{1}p^{2m_2},\ \quad \rho_{34}=\mu^2N^{2}_{1}N'_1p^{m_2},\ \quad \rho_{44}=\mu^2N^{2}_{1}N'^{2}_{1}.\\
 \end{array}
\ee
The matrix  $R$, associated to   $\rho_{AB}$, has two nonzero   eigenvalues $\lambda_1, \lambda_2$ (i.e. $\lambda_3=\lambda_4=0$)    where  $\lambda_1>\lambda_2$ if $\mu<0$ and $\lambda_2>\lambda_1$ if $\mu>0$, whence
\be
      C_{AB}=\max\{0,|\lambda_1-\lambda_2|\} \quad \mathrm{\ for\ all}\ \ \mu.
\ee
The concurrence $C_{AB}$  can be represented diagrammatically as a function of $0\leq p<1$ and $\mu$ as shown in the figure \ref{cab}.
For negative $\mu$ the maximum value $C_{AB}$ occurs at $\mu=-1$ when $p\rightarrow 1$,   while for positive $\mu$
the maximum occurs  at $\mu=1$  and a point $p<1$.
 Similar consequences  hold for concurrence  $C_{AD}$  (see figure \ref{profile1}).
On the other hand, if we define $\tau_{ABD}=C_{A(BD)}^2-C_{AB}^2-C_{AD}^2$ then the positivity of $\tau_{ABD}$ examine the monogamy inequality (\ref{monogamy}) (see figure \ref{mono1}).
Figure \ref{mono1} shows that the difference between the left and right-hand side of inequality (\ref{monogamy}) decrease by increasing $p$ (see also figure \ref{profile2}).
Later, in the next section we will revisit the monogamy inequality for qutrit states.
\section{Multi-qutrit case}
\begin{figure}
\centerline{\includegraphics[width=8cm]{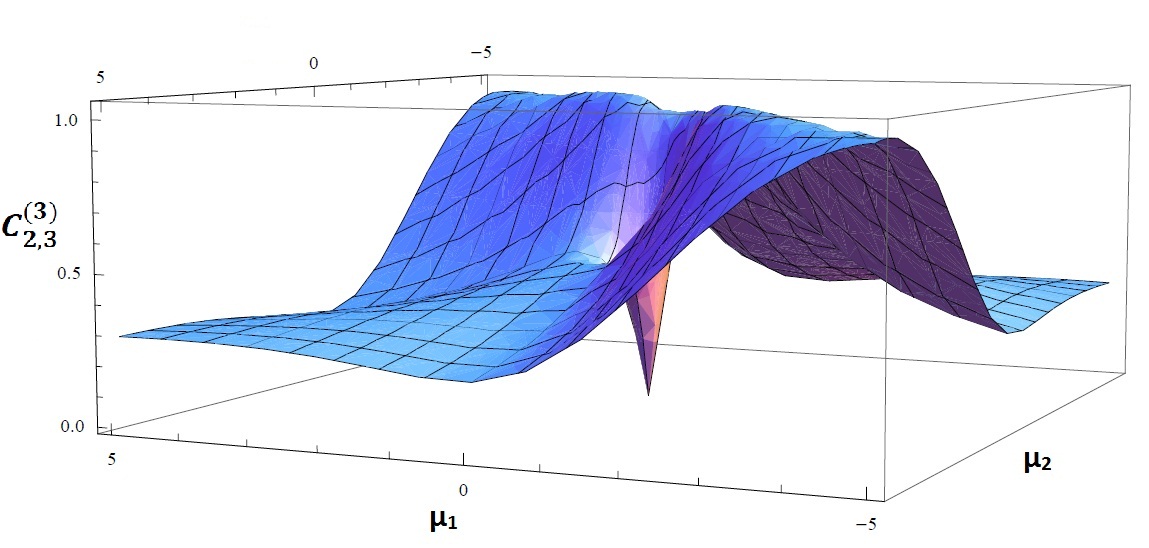}}
  \caption{\small Concurrence $C^{(3)}_{2,3}$ as a function of $\mu_{1}$ and $\mu_{2}$ for $p_{1}=0.9$, $p_{2}=0.89$ and $p_{3}=0.8$.}\label{qutrit}
\end{figure}

\begin{figure}
\centerline{\includegraphics[width=14cm]{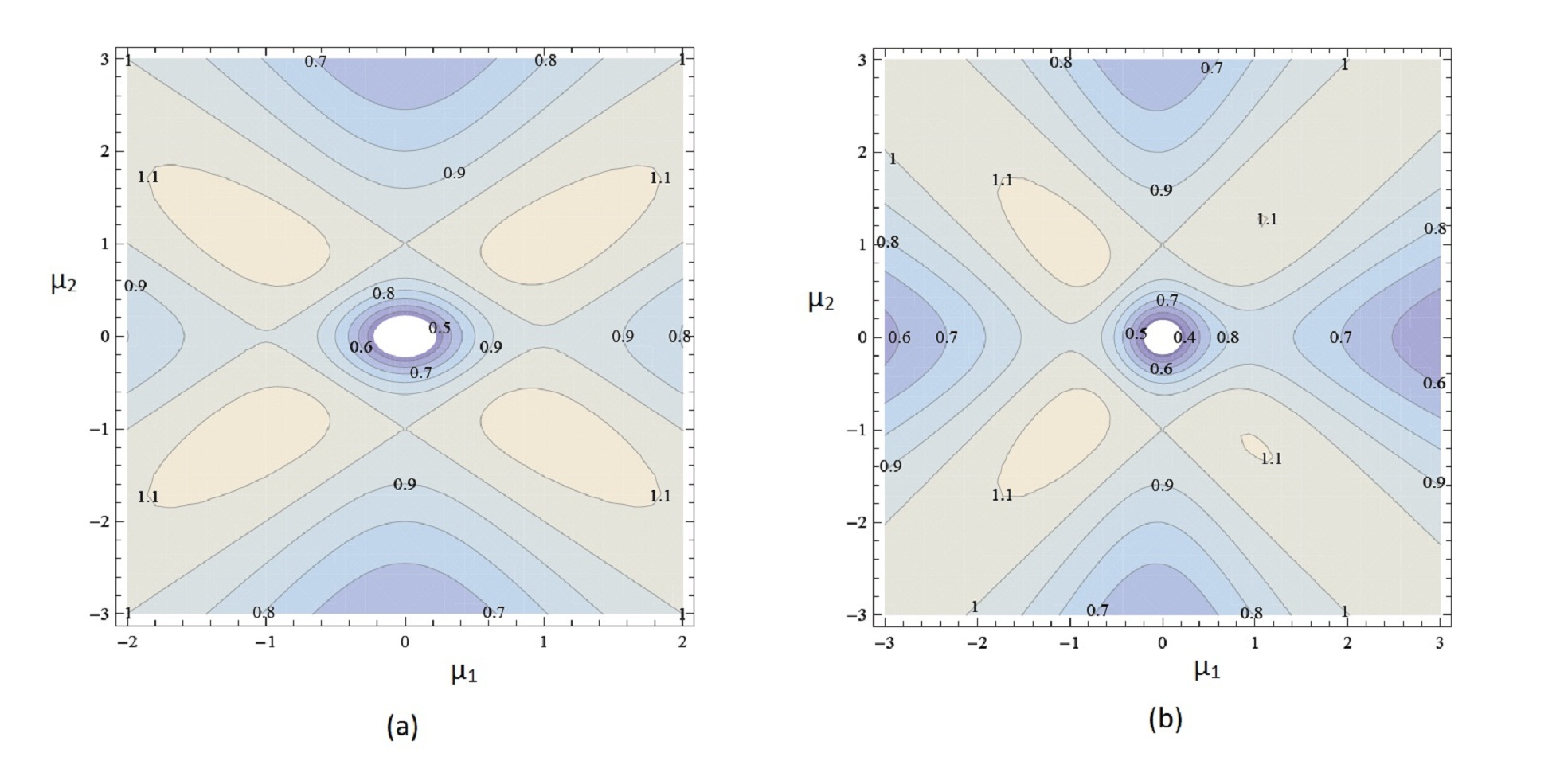}}
  \caption{\small Concurrence $C^{(3)}_{2,3}$ as a function of $\mu_{1}$ and $\mu_{2}$ for (a): $p_{1}=0.3$  and
 (b):  $p_{2}=0.6$.}\label{profile3}
\end{figure}

It is tempting to add further terms to state (\ref{4}) and explore its  behavior  from entanglement point of view.
To this end, let us consider the   N-mode state
\be\label{si1}
|\Psi^{(3)}_{N}\rangle=\frac{1}{\sqrt{M^{(3)}_N}}(|\alpha\rangle\cdots|\alpha\rangle+\mu_{1}|\beta\rangle\cdots|\beta\rangle+\mu_{2}|\gamma\rangle\cdots|\gamma\rangle),
\ee
 where $p_{1}=\langle\alpha|\beta\rangle$,  $p_{2}=\langle\gamma|\beta\rangle$, $p_{3}=\langle\gamma|\alpha\rangle$
(for simplicity we assume that the  all parameters  are real) and
\be
M^{(3)}_N=1+\mu^{2}_1+\mu^{2}_2+2\mu_{1}p^{N}_1+2\mu_{2}p^{N}_3+2\mu_{1}\mu_{2}p^{N}_2.
\ee
We again assume that the three non-orthogonal coherent states $|\alpha\rangle, |\beta\rangle$ and
 $|\gamma\rangle$ are  linearly independent and span a three-dimensional subspace of the Hilbert space. Therefore  we can define
three orthonormal basis as
\be\label{newbase}
\begin{array}{l}
|0\rangle=|\alpha\rangle,\\
|1\rangle=\frac{1}{\sqrt{1-p_{1}^{2}}}(|\beta\rangle-p_{1}|\alpha\rangle),\\
|2\rangle=\sqrt{\frac{1-p_{1}^{2}}{1-p^2_{1}-p^2_{2}-p^2_{3}+2p_1p_2p_3}}\left(|\gamma\rangle+(\frac{p_{1}p_{3}-p_{2}}{{1-p_{1}^{2}}})|\beta\rangle+(\frac{p_{1}p_{2}-p_{3}}{{1-p_{1}^{2}}})|\alpha\rangle\right).
\end{array}
\ee
Once again, we can assume  the first $m\ (\leq\frac{N}{2})$ modes as part one and the  remaining  $(N-m)$ modes as the second part and rewrite
(\ref{si1}) as
\be\label{phi}
 \begin{array}{l}
 |\Psi^{(3)}_{N}\rangle=\frac{1}{\sqrt{M^{(3)}_{N}}}\left( (1+\mu_1 p^{N}_1+\mu_2 p^{N}_3)|\mathbf{00'}\rangle+(\mu_1N_1p^{N-m}_{1}-\mu_2xN_1p^{N-m}_{3})|\mathbf{10'}\rangle\right.\\
\left. ~~~~~~~+(\mu_1N'_1p^{m}_{1}-\mu_2x'N'_1p^{m}_{3})|\mathbf{01'}\rangle+(\mu_1N_1N'_{1}-\mu_2xx'N_1N'_1)|\mathbf{11'}\rangle-\mu_2 x'N'_{1}N_{2}|\mathbf{21'}\rangle\right. \\
 \left.~~~~~~~-\mu_2xN_{1}N'_{2}|\mathbf{12'}\rangle+
 \mu_2N_{2}p^{N-m}_{3}|{\mathbf{20}'}\rangle+\mu_2N'_{2}p^{m}_{3}|\mathbf{02'}\rangle+\mu_2N_{2}N'_{2}|\mathbf{22'}\rangle\right) ,
 \end{array}
 \ee
in which the new basis is defined as usual qutrit like basis
 \be
\begin{array}{l}
|\mathbf{0}\rangle\equiv\underbrace{|0\rangle\cdots|0\rangle}_{m},\quad\quad|\mathbf{0'}\rangle\equiv\underbrace{|0\rangle\cdots|0\rangle}_{N-m},\\
|\textbf{1}\rangle\equiv\underbrace{|1\rangle\cdots|1\rangle}_{m},\quad\quad|\mathbf{1'}\rangle\equiv\underbrace{|1\rangle\cdots|1\rangle}_{N-m},\\
|\textbf{2}\rangle\equiv\underbrace{|2\rangle\cdots|2\rangle}_{m},\quad\quad|\mathbf{2'}\rangle\equiv\underbrace{|2\rangle\cdots|2\rangle}_{N-m},
\end{array}
\ee
and
 \be
 \begin{array}{l}
 x=\frac{p_{1}^{m}p_{3}^{m}-p_{2}^{m}}{{1-p_{1}^{2m}}},\quad\quad x'=\frac{p_{1}^{N-m}p_{3}^{N-m}-p_{2}^{N-m}}{{1-p_{1}^{2(N-m)}}}, \\ y=\frac{p_{1}^{m}p_{2}^{m}-p_{3}^{m}}{{1-p_{1}^{2m}}},\quad\quad y'=\frac{p_{1}^{N-m}p_{2}^{N-m}-p_{3}^{N-m}}{{1-p_{1}^{2(N-m)}}},\\
 N_{1}=\sqrt{1-p_{1}^{2m}}, N_{1}'=\sqrt{1-p_{1}^{2(N-m)}},\\
 N_{2}=\sqrt{1-p_{3}^{2m}-x^{2}N_{1}^{2}}$, $N_{2}'=\sqrt{1-p_{3}^{2(N-m)}-x'^{2}N_{1}'^{2}},\\
 \end{array}
 \ee
hence the $|\Psi^{(3)}_{N}\rangle$ recast as two qutrit state. It is worthwhile to emphasize that the three parameters  $p_{i}$ (i=1,2,3) are not, in general, independent of each other. This comes from the fact that both $N^2_{2}$ and $ N'^2_{2}$, being normalization factors, must be positive definite which impose the following constrains
\be
 \left\{\begin{array}{l}
\lambda_{-}<p^{N-m}_{3}<\lambda_{+},\quad \mathrm{if}\quad\quad p^{2(N-m)}_{1}+p^{2(N-m)}_{2}\geq1,\\
0<p^{N-m}_{3}<\lambda_{+},\quad\quad\mathrm{if}\quad\quad p^{2(N-m)}_{1}+p^{2(N-m)}_{2}<1,
 \end{array}\right.
\ee
where
\be
 \lambda_{\pm}=(p_{1}p_{2})^{N-m}\pm\sqrt{1-p^{2(N-m)}_{1}-p^{2(N-m)}_{2}+(p_{1}p_{2})^{2(N-m)}}.
 \ee
 We can now use the concurrence formulae (\ref{cc}) to evaluate the entanglement of bipartite  state $|\varphi_{N}\rangle$, i.e.
 \be\label{bi}
 \begin{array}{l}
 C^{(3)}_{m,N-m}=\frac{1}{M^{(3)}_{N}}\left( |N_{1}N_{1}'(\mu_{1}+\mu_{2}xx')+\mu_{1}\mu_{2}N_{1}N_{1}'(xx'p_{1}^{N}+p_{3}^{N}+x'p_{3}^{m}p_{1}^{N-m}+xp_{3}^{N-m}p_{1}^{m})|^2\right.\\
 \left.~~~+|N_{1}N_{2}'(\mu_{2}x+\mu_{1}\mu_{2}xp_{1}^{N}+\mu_{1}\mu_{2}p_{3}^{m}p_{1}^{N-m})|^2+|N_{1}'N_{2}(\mu_{2}x'+\mu_{1}\mu_{2}x'p_{1}^{N}+\mu_{1}\mu_{2}p_{3}^{N-m}p_{1}^{m})|^2\right.\\
 \left.~~~+|\mu_{2}N_{2}N_{2}'(1+\mu_{1}p_{1}^{N})|^2+|\mu_{1}\mu_{2}N_{1}N_{1}'N_{2}'y|^2+|\mu_{1}\mu_{2}N_{1}'N_{2}N_{2}'p_{1}^{m}|^2
+|\mu_{1}\mu_{2}N_{1}N_{1}'N_{2}y'|^2\right.\\
\left. ~~~+|\mu_{1}\mu_{2}N_{1}N_{2}N_{2}'p_{1}^{N-m}|^2+|\mu_{1}\mu_{2}N_{1}N_{1}'N_{2}N_{2}'|^2\right){^{\frac{1}{2}}}.\\
\end{array}
\ee
The concurrence $C^{(3)}_{m,N-m}$ is reduced to $C^{(2)}_{m,N-m}$, when one of the $\mu_{1}$ or $\mu_{2}$ become zero.
 Let us first determine the separable states, i.e. $C^{(3)}_{m,N-m}=0$ which entail vanishing of  all  absolute terms appearing in $C^{(3)}_{m,N-m}$. Using the vanishing of  final term implies that $\mu_{1}$ or $\mu_{2}$ must be zero. The forth term impose that $\mu_{2}=0$ and subsequently the first term demands that $\mu_1$ must  also be zero. Hence the state $|\Psi^{(3)}_{N}\rangle$ is bipartite separable, if and only if both $\mu_{1}$ and $\mu_{2}$ vanish. On the other hand, the  state (\ref{phi}) has maximum concurrence,  i.e. $(C^{(3)}_{m,N-m})_{max}=\sqrt{\frac{4}{3}}$ if  $p_i=0$ and  $\mu_{1,2}=\pm1$ which means that  it  reduces to GHZ like states $|\Psi^{(3)}_{N}\rangle_{GHZ}=\frac{1}{\sqrt{3}}(|\mathbf{00'}\rangle\pm)|\mathbf{11'}\rangle\pm|\mathbf{22'}\rangle)$.
 We illustrate the behaviour of $C^{(3)}_{2,3}$ as a function of $\mu_{1}$ and $\mu_{2}$ for given values $p_{i}$ in figure \ref{qutrit}.
We note that the concurrence $C^{(3)}_{2,3}$ is  sensitive for the different  values  of the parameters. To see this, one may take $p_2=p_3=0$, (i.e. $\gamma\rightarrow\infty$) and $p_1\neq0$ that leads to  the state
\be
 \begin{array}{l}
 |\Psi^{(3)}_{5}\rangle=\frac{1}{\sqrt{1+\mu^2_1+\mu^2_2+2\mu_1p^{5}_1}}\left( (1+\mu_1 p^{5}_1)|\mathbf{00'}\rangle+\mu_1N_1p^{3}_{1}|\mathbf{10'}\rangle\right.\\
\left. ~~~~~~~+\mu_1N'_1p^{2}_{1}|\mathbf{01'}\rangle+\mu_1N_1N'_{1}|\mathbf{11'}\rangle+\mu_2|\mathbf{22'}\rangle\right) ,
 \end{array}
 \ee
with the following   concurrence
\be
 \begin{array}{l}
 C^{(3)}_{2,3}=\frac{1}{1+\mu^2_1+\mu^2_2+2\mu_1p^{5}_1}\left( |\mu_{1}N_{1}N_{1}'|^2
+|\mu_{2}(1+\mu_{1}p_{1}^{5})|^2\right.\\
\left.\quad\quad\quad\quad+|\mu_{1}\mu_{2}N_{1}'p_{1}^{2}|^2+|\mu_{1}\mu_{2}N_{1}p_{1}^{3}|^2+|\mu_{1}\mu_{2}N_{1}N_{1}'|^2\right){^{\frac{1}{2}}}.\\
\end{array}
\ee
The behaviour of concurrence  $C^{(3)}_{2,3}$ as a function of  $\mu_1$ and $\mu_2$ for given $p_1$ is shown in figure \ref{profile3}.
\subsection{Mixed states and monogamy inequality for multi-qutrit case}
\begin{figure}
\centerline{\includegraphics[width=10cm]{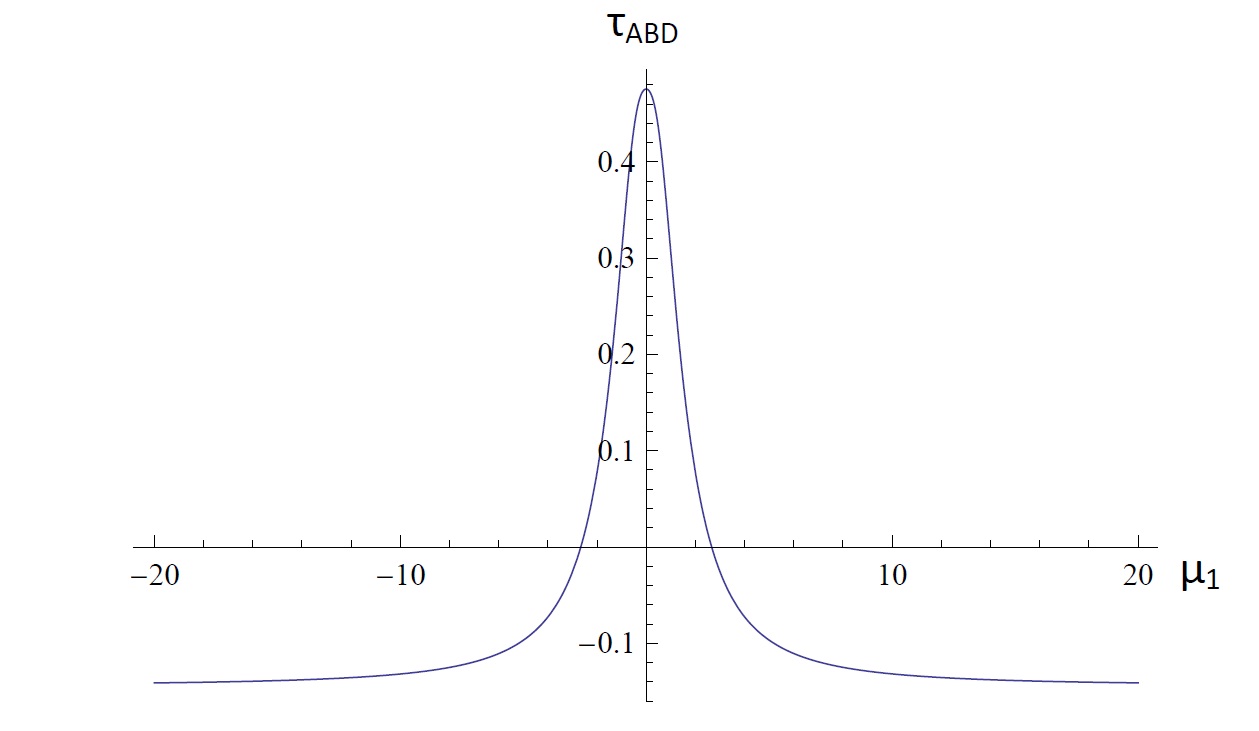}}
  \caption{\small $\tau_{ABD}$ as a function of $\mu_1$ for given $N=20, m_1=1, m_2=2, p_1=0.7$ and $\mu_2=0.4$. For $\mu_1>2.588$ we have a violation of monogamy inequality (\ref{monogamy}).}\label{mo}
\end{figure}

Now let us partition the state (\ref{si1}) to tripartite A,B and D including $m_1$, $m_2$ and $m_3=N-m_1-m_2$ modes respectively.
For the moment, suppose that $\gamma\rightarrow\infty$, i.e. $p_2,p_3=0$ and $p_1\neq0$. Then  the state (\ref{si1}) is reduced to three-qutrit state as
\be
\begin{array}{l}
|\Psi^{(3)}_{N}\rangle_{ABD}=\frac{1}{\sqrt{M^{(3)}_N}}((1+\mu_1p^N_1)|\mathbf{0}\mathbf{0'}\mathbf{0''}\rangle+\mu_1N''_1p^{n+m}_1|\mathbf{0}\mathbf{0'}\mathbf{1''}\rangle+\mu_1N'_1p^{N-m}_1|\mathbf{0}\mathbf{1'}\mathbf{0''}\rangle\\
\quad\quad\quad\quad+ \mu_1N_1p^{N-n}_1|\mathbf{1}\mathbf{0'}\mathbf{0''}\rangle+\mu_1N'_1N''_1p^{n}_1|\mathbf{0}\mathbf{1'}\mathbf{1''}\rangle+\mu_1N_1N''_1p^{m}_1|\mathbf{1}\mathbf{0'}\mathbf{1''}\rangle\\
\quad\quad\quad\quad+\mu_1N_1N'_1p^{N-m-n}_1|\mathbf{1}\mathbf{1'}\mathbf{0''}\rangle+\mu_1N_1N'_1N''_1|\mathbf{1}\mathbf{1'}\mathbf{1''}\rangle+\mu_2|\mathbf{2}\mathbf{2'}\mathbf{2''}\rangle),
\end{array}
\ee
where $N_{1}=\sqrt{1-p_{1}^{2m_{1}}}$, $N'_{1}=\sqrt{1-p_{1}^{2m_{2}}}$, $N''_{1}=\sqrt{1-p_{1}^{2 m_{3}}}$ and
\be
\begin{array}{l}
|\mathbf{0}\rangle\equiv\underbrace{|0\rangle\cdots|0\rangle}_{m_{1}},\quad\quad|\mathbf{0}'\rangle\equiv\underbrace{|0\rangle\cdots|0\rangle}_{m_{2}},\quad\quad|\mathbf{0}''\rangle\equiv\underbrace{|0\rangle\cdots|0\rangle}_{m_{3}},\\
|\mathbf{1}\rangle\equiv\underbrace{|1\rangle\cdots|1\rangle}_{m_{1}},\quad\quad|\mathbf{1}'\rangle\equiv\underbrace{|1\rangle\cdots|1\rangle}_{m_{2}},\quad\quad|\mathbf{1}''\rangle\equiv\underbrace{|1\rangle\cdots|1\rangle}_{m_{3}},\\
|\mathbf{2}\rangle\equiv\underbrace{|2\rangle\cdots|1\rangle}_{m_{1}},\quad\quad|\mathbf{2}'\rangle\equiv\underbrace{|2\rangle\cdots|2\rangle}_{m_{2}},\quad\quad|\mathbf{2}''\rangle\equiv\underbrace{|2\rangle\cdots|2\rangle}_{m_{3}}.
\end{array}
\ee
One  can easily obtain the reduced density matrices $\rho_{AB}=Tr_{_{D}}(|\Psi^{(3)}_{N}\rangle_{ABD}\langle\Psi^{(3)}_{N}|)$ and $\rho_{AD}=Tr_{_{B}}(|\Psi^{(3)}_{N}\rangle_{ABD}\langle\Psi^{(3)}_{N}|)$ by partially tracing out subsystems $D$ and $B$ respectively.  For general mixed states the concurrence is defined as $|C|^2=\sum_{\alpha\beta}|C_{\alpha\beta}|^2$ where $C^{\alpha\beta}=\lambda^{\alpha\beta}_1-\sum^n_{i=2}\lambda^{\alpha\beta}_i$ with $\lambda_1=\max\{\lambda_i, i=1,...,d^2\}$ and $\lambda^{\alpha\beta}_i$ are the nonnegative square root of eigenvalues of  $\tau\tau^*$ defined as \cite{YOU}
\be\label{yo}
\tau^{\alpha\beta}\tau^{\alpha\beta*}=\sqrt{\rho}(E_\alpha-E_{-\alpha})\otimes(E_\beta-E_{-\beta})\rho^*(E_\alpha-E_{-\alpha})\otimes(E_\beta-E_{-\beta})\sqrt{\rho}.
\ee
If we take  $N=20, m_1=1, m_2=2, p=0.7$ and $\mu_2=0.4$, then the components of concurrence vectors $C_{AB}$ and $C_{AD}$,  for all $\mu_1$, read
\be
\begin{array}{l}
{C}_{AB}=(\frac{|\mu_1|(0.0034+4.6\times10^{-6}\mu_1+0.003\mu^2_1)}{1.35+0.004\mu_1+2.32\mu^2_1+0.0032\mu^3_1+\mu^4_1},0,0,0,0,0,\frac{0.38\mu^2_1}{1.16+0.002\mu_1+\mu^2_1},0,0),\\
C_{AD}=(\frac{|\mu_1|(0.81+0.001\mu_1+0.7\mu^2_1)}{1.35+0.004\mu_1+2.32\mu^2_1+0.0032\mu^3_1+\mu^4_1},0,0,0,0,0,-\frac{0.35\mu_1}{1.16+0.002\mu_1+\mu^2_1},0,0).
\end{array}
\ee
On the other hand,  using (\ref{bi}), one finds that
\be
C^2_{A(BD)}=\frac{0.64+0.001\mu_1+1.54\mu^2_1}{1.16+0.002\mu_1+\mu^2_1}.
\ee
The behaviour of $\tau_{ABD}=C_{A(BD)}^{2}-C^2_{AB}-C^2_{AD}$ as a function of $\mu_1$ is shown in figure \ref{mo}.
One can immediately deduce that $\tau_{ABD}$ becomes negative for $\mu_1>2.588$ which is a violation of the inequality in Eq. (\ref{monogamy}). Clearly, for $\mu_1<2.588$, the monogamy inequality is satisfied.
The other example which fulfils the monogamy inequality is the state
$\frac{1}{\sqrt{M}}(|\mathbf{0}\mathbf{0'}\mathbf{0''}\rangle+\mu_1|\mathbf{1}\mathbf{1'}\mathbf{1''}\rangle+\mu_2|\mathbf{2}\mathbf{2'}\mathbf{2''}\rangle)$
for which we have $C_{AB}=C_{AD}=0$ and $C_{A(BD)}=\frac{2\sqrt{\mu^2_1+\mu^2_2+\mu^2_1\mu^2_2}}{1+\mu^2_1+\mu^2_2}\geq0$, hence $\tau_{ABD}\geq0$ for all $\mu_1$ and $\mu_2$. The extremum inequality is accomplished for $\mu_1,\mu_2=\pm1$, namely $C_{A(BD)}^{2}=4/3\geq 0=C_{AB}^{2}+C_{AD}^{2}$.
\section{Separability of balanced N-mode  coherent state}
For completeness we mention that the separability of the aforementioned  states, discussed above, refer to vanishing of all coefficients $\mu_{i}$,  does not restrict to   qubit and qutrit cases. To see this consider the following general balanced N-mode coherent state
\be\label{sid}
|\Psi^{(d)}_{N}\rangle=\frac{1}{\sqrt{M^{(d)}_N}}(\mu_0|\alpha_0\rangle\cdots|\alpha_0\rangle+
\mu_1|\alpha_1\rangle\cdots|\alpha_1\rangle+\cdots+\mu_{d-1}|\alpha_{d-1}\rangle\cdots|\alpha_{d-1}\rangle),
\ee
where without loss of generality we can take $\mu_0=1$.
We wish to show that the above state is bipartite separable if and only if all  $\mu_i=0, \ \mathrm{with}\ i=1,2,...,d-1$. Once again, to keep our discussion simple (yet generic in scope), let us focus attention on  $d=3,4$  cases.
\par
\textbf{Case $d=3$}: For qutrit states
 the concurrence is $C^{(3)}_{m,N-m} = 2  ({\sum\limits_{i < j}^{2 } {\sum\limits_{k < l}^{2 } {\left| {a_{ik} a_{jl}  - a_{il} a_{jk} } \right|^2 } })^\frac{1}{2} }$. The  term $|a_{11}a_{22}-a_{12}a_{21}|$ is equal to last term $|\mu_{1}\mu_{2}N_{1}N_{1}'N_{2}N_{2}'|$ in Eq. (\ref{bi}) which contains the product $\mu_1\mu_2$. Therefore, we have two cases: $\mu_1=0,\mu_2\neq0$ or $\mu_1\neq0, \mu_2=0$, (the case both $\mu_{1}=\mu_{2}=0$ is trivial).
 The first case contradicts the vanishing of the term $|a_{00}a_{22}-a_{02}a_{20}|=|\mu_{2}N_{2}N_{2}'(1+\mu_{1}p_{1}^{N})|$, while the second case  in contrast to vanishing of $|a_{00}a_{11}-a_{01}a_{10}|=|N_{1}N_{1}'(\mu_{1}+\mu_{2}xx')+\mu_{1}\mu_{2}N_{1}N_{1}'(xx'p_{1}^{N}+p_{3}^{N}+x'p_{3}^{m}p_{1}^{N-m}+xp_{3}^{N-m}p_{1}^{m})|$.
Hence, the qutrit state $|\Psi^{(3)}_{N}\rangle$ is bi-separable if and only if both $\mu_{1}=\mu_{2}=0$.
\par
\textbf{Case $d=4$}: One step further  is to consider  the following state
\be\label{si2}
|\Psi^{(4)}_{N}\rangle=\frac{1}{\sqrt{M^{(4)}_N}}|\alpha_0\rangle\cdots|\alpha_0\rangle
+\mu_{1}|\alpha_1\rangle\cdots|\alpha_1\rangle+\mu_{2}|\alpha_2\rangle\cdots|\alpha_2\rangle
+\mu_{3}|\alpha_3\rangle\cdots|\alpha_3\rangle.
\ee
The last term of concurrence $C^{(4)}_{m,N-m} = 2  ({\sum\limits_{i < j}^{3 } {\sum\limits_{k < l}^{3 } {\left| {a_{ik} a_{jl}  - a_{il} a_{jk} } \right|^2 } })^\frac{1}{2} }$, i.e. $|a_{22}a_{33}-a_{23}a_{32}|\propto\mu_2\mu_3$ which is equal zero for all $p_i$ if  $\mu_2=0,\mu_3\neq0$ or $\mu_2\neq0, \mu_3=0$. The former case contradicts the  vanishing of the terms $|a_{11}a_{33}-a_{13}a_{31}|\propto\mu_3(\mu_1+\mu_2xx')$ and $|a_{00}a_{33}-a_{03}a_{30}|\propto\mu_3(1+\mu_1p^N_1+\mu_2p^N_3)$, while the latter case  contradicts the the vanishing of the terms
\be
\begin{array}{c}
  |a_{11}a_{22}-a_{12}a_{21}|\propto \mu_1(\mu_2+f_1\mu_3)+f_2\mu_2\mu_3,\\
  |a_{00}a_{22}-a_{02}a_{20}|\propto \mu_2+f_3\mu_3+f_4\mu_1\mu_2+f_5\mu_1\mu_3+f_6\mu_2\mu_3.
\end{array}
\ee
 Hence both  $\mu_2,\mu_3=0$. On the other hand, the vanishing of  the first term
\be
  |a_{00}a_{11}-a_{01}a_{10}|\propto \mu_1+g_1\mu_2+g_2\mu_3+g_3\mu_1\mu_2+g_4\mu_1\mu_3+g_5\mu_2\mu_3,
\ee
imposes that $\mu_1=0$, where $f_i$'s and $g_i$'s are some  functions of $p_i$'s.
\par
\textbf{General case} :
We next turn to the general discussion of the problem with arbitrary $d$. As should be clear from the discussion above,
in each step the vanishing of the term $|a_{ii}a_{_{jj}}-a_{_{ij}}a_{_{ji}}|, i<j$  leads to two cases $\mu_i=0,\mu_j\neq0$ or $\mu_i\neq0, \mu_j=0$. The former case contradicts the vanishing of the terms $|a_{ii}a_{_{j'j'}}-a_{_{ij'}}a_{_{j'i}}|, j'=0,1,...,j-2$, while the latter case contradicts the vanishing of the terms $|a_{i'i'}a_{_{jj}}-a_{_{i'j}}a_{_{ji'}}|, i'=0,1,...,i-1$.
\section{Generation of balanced N-mode  entangled coherent state: Even terms}
In this section, we want to propose a scheme to produce the general balanced N-mode entangled coherent states. To this aim we first  need to have  the superposition of even and odd number of coherent states like as $|\alpha\rangle+\mu|\beta\rangle$ and $|\alpha\rangle+\mu_1|\beta\rangle+\mu_2|\gamma\rangle$, (up to normalization factors), and so on. In general, the superpositions of coherent states are difficult to produce, and fundamentally this could be due to extreme sensitivity to environmental decoherence (see, e.g., \cite{Cheong1,Enk1, Sanders3,Milburn1,Milburn2,Yurke1,Yurke2}). Here we restrict ourselves  to the even cases.
 One may  use the displacement operator together with parity operation \cite{mesina1,mesina2}, to construct the unitary operation
\be
\hat{U}(\lambda,\alpha)=e^{i\lambda{\hat{D}}(\alpha)\hat{\Pi}},
\ee
where $\lambda$ is real number, $\hat{\Pi}=\cos(\pi \hat{a}^{\dag} \hat{a})$ is a Hermitian and unitary operator with property $\hat{\Pi}|n\rangle=(-1)^{n}|n\rangle$, and $\hat{D}(\alpha)$ is usual displacement operator (\ref{displacement}).
Using the fact that $[\hat{D}(\alpha)\hat{\Pi}]^{2}=\hat{I}$, it can be rewrite  as
\be
\hat{U}(\lambda,\alpha)=\cos\lambda\hat{I}+i\sin\lambda\hat{D}(\alpha)\hat{\Pi}.
\ee
The action of such an operator on a vacuum state is
\be
\hat{U}(\lambda,\alpha)|0\rangle=\cos\lambda|0\rangle+i\sin\lambda|\alpha\rangle.
\ee
In order to obtain a linear combination of two arbitrary Glauber coherent states, it is enough  to use another displacement operator, $\hat{D}(\beta)$:
\be
\hat{D}(\beta)\hat{U}(\lambda,\alpha)|0\rangle=\cos\lambda|\beta\rangle+i\sin\lambda|\alpha+\beta\rangle.
\ee
Using $\hat{V}(\alpha,\beta,\lambda)=\hat{D}(\alpha)\hat{U}(\lambda,\beta-\alpha)$, one may recast the above state in a convenient form as follows
\be\label{1}
\hat{V}(\alpha,\beta,\lambda)|0\rangle=\cos\lambda|\alpha\rangle+i\sin\lambda e^{iIm(\alpha\beta^{*})}|\beta\rangle.
\ee
 The method reported here may be extended to the case of a superposition involving just $2^{N}$ coherent states, by considering the following multiplication
\be
\hat{V}^{N}=\prod_{k=1}^{N}\hat{V}(\lambda_{k},\alpha_{k},\beta_{k}).
\ee
We  next  use  polarizing beam splitter (PBS). The polarizing beam splitter is commonly made by cementing together two birefringent materials like calcite or quartz, and has the property of splitting a light beam into its orthogonal linear polarizations.
The beam splitter interaction given by the unitary transformation
\be
\hat{B}_{i-1,i}(\theta)=\exp[\theta(\hat{a}_{i-1}^{\dag}\hat{a}_{i}-\hat{a}_{i}^{\dag}\hat{a}_{i-1})].
\ee
which $\hat{a}_{i-1}$, $\hat{a}_{i}$, $\hat{a}^{\dag}_{i-1}$ and $\hat{a}^{\dag}_{i}$ are the annihilation and creation operators of the field mode $i-1$ and $i$, respectively. Using Baker-Hausdorf formula, the action of the beam splitter on two modes $i-1$ and $i$ ,  can be expressed as
\be\label{action}
\hat{B}_{i-1,i}(\theta)\left(
  \begin{array}{c}
    \hat{a}_{i-1} \\
    \hat{a}_{i} \\
  \end{array}
\right)\hat{B}^{\dag}_{i-1,i}(\theta)=\left(
  \begin{array}{c}
    \hat{a}'_{i-1} \\
    \hat{a}'_{i} \\
  \end{array}
\right)=\left(
          \begin{array}{cc}
            \cos\theta & -\sin\theta \\
            \sin\theta & \cos\theta \\
          \end{array}
        \right)\left(
                 \begin{array}{c}
                   \hat{a}_{i-1} \\
                   \hat{a}_{i} \\
                 \end{array}
               \right).
\ee
We consider $50-50$ beam splitter, i.e. $\theta=\pi/4$ with the following operation
 \be
\hat{B}_{1,2}(\pi/4)|\alpha'\rangle_{_{1}}|0\rangle_{_{2}}=|\frac{\alpha'}{\sqrt{2}}\rangle_{_{1}}|\frac{\alpha'}{\sqrt{2}}\rangle_{_{2}}.
\ee
This  result says that like classical light wave where the incident intensity is evenly divided between the two output beams, e.g. half the incident average photon number, $\frac{|\alpha|^{2}}{2}$, emerges in each beam. Note that the output is not entangled. For producing entangled coherent state suppose that our input state be a superposition of two coherent states as $|\alpha'\rangle_{_{1}}+\mu_1|\beta'\rangle_{_{1}}$.  Following the procedure above, we may then, obtain the output state as
\be
\hat{B}_{1,2}(\pi/4)(|\alpha'\rangle_{_{1}}+\mu_1|\beta'\rangle_{_{1}})\otimes|0\rangle_{_{2}}=|\frac{\alpha'}{\sqrt{2}}\rangle_{_{1}}|\frac{\alpha'}{\sqrt{2}}\rangle_{_{2}}+
\mu_1|\frac{\beta'}{\sqrt{2}}\rangle_{_{1}}|\frac{\beta'}{\sqrt{2}}\rangle_{_{2}}.
\ee
By renaming $\alpha\equiv\frac{\alpha'}{\sqrt{2}}$ and $\beta\equiv\frac{\beta'}{\sqrt{2}}$ we have
$
|\alpha\rangle_{_{1}}|\alpha\rangle_{_{2}}+\mu_1|\beta\rangle_{_{1}}|\beta\rangle_{_{2}},
$
which is a two-mode entangled coherent state.
In order to obtain the three-mode entangled coherent states we use two  beam splitters with reflectivity amplitude of $\frac{1}{\sqrt{3}}$, modes 1 and 2.
 At  first beam splitter, the modes 1 and 2 are combined with reflectivity amplitude of $\frac{1}{\sqrt{3}}$ and subsequently, the modes 2 and 3 undergo  a 50-50 beam splitter, i.e.
\be
\begin{array}{c}
  \hat{B}_{2,3}(\pi/4)\hat{B}_{1,2}(\cos^{-1}(\frac{1}{\sqrt{3}}))[(|\alpha'\rangle_{_{1}}+\mu_1|\beta'\rangle_{_{1}})\otimes|0\rangle_{_{2}}\otimes|0\rangle_{_{3}}] \\
  =|\frac{\alpha'}{\sqrt{3}}\rangle_{_{1}}|\frac{\alpha'}{\sqrt{3}}\rangle_{_{2}}|\frac{\alpha'}{\sqrt{3}}\rangle_{_{3}}+
\mu_1|\frac{\beta'}{\sqrt{3}}\rangle_{_{1}}|\frac{\beta'}{\sqrt{3}}\rangle_{_{2}}|\frac{\beta'}{\sqrt{3}}\rangle_{_{3}},
\end{array}
\ee
which can be recasted in the $|\alpha\rangle_{_{1}}|\alpha\rangle_{_{2}}|\alpha\rangle_{_{3}}+\mu_1|\beta\rangle_{_{1}}|\beta\rangle_{_{2}}|\beta\rangle_{_{3}}.$
The procedure  can be easily generalized to prepare N-mode $(N=2^k)$ entangled coherent states. To this end,  we should first apply the unitary operator  $\hat{V}^{2^k}=\prod_{i=1}^{{2^k}}\hat{V}(\lambda_{i},\alpha_{i},\beta_{i})$ to the  initial state $|0\rangle_{_{1}}$ and subsequently use the sequential beam splitters  $\hat{B}_{N-1,N}\hat{B}_{N-2,N-1}...\hat{B}_{2,3}\hat{B}_{1,2}$
with $\frac{1}{\sqrt{2}}, \frac{1}{\sqrt{3}},...,\frac{1}{\sqrt{N-1}}, \frac{1}{\sqrt{N}}$ reflectivity amplitude respectively. The final result is a qudit like $2^k$-mode entangled coherent state (\ref{mainstate}) (see figure 13).
\begin{figure}
\centerline{\includegraphics[width=15cm]{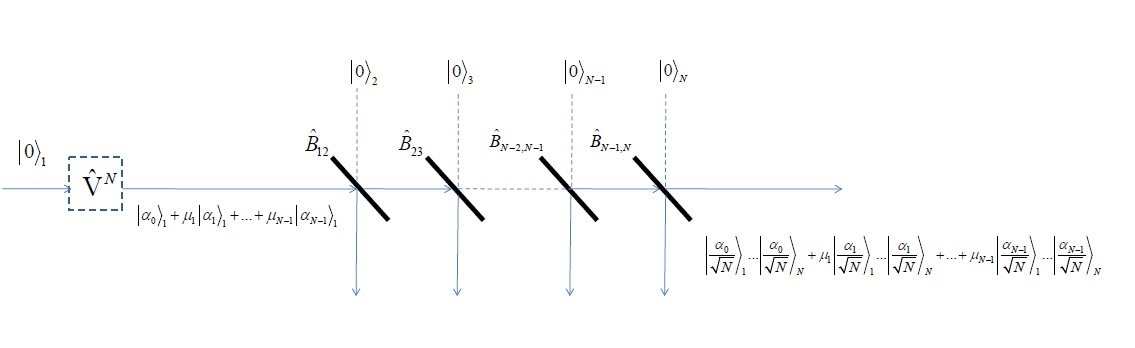}}
  \caption{\small A protocol for generation of balanced N-mode  entangled coherent states}\label{3}
\end{figure}
The scheme to generate N-mode entangled coherent state with N odd is under debate. As mentioned above, the problem is to produce the superposition of arbitrary coherent states. For example a cat state  $|\alpha\rangle_{\pm}=|\alpha\rangle\pm|-\alpha\rangle$,
served as a qubit for quantum logical encoding, can be easily produce in a Kerr medium \cite{Milburn} while the problem of generating discrete superpositions of coherent states in the process of light propagation through a nonlinear Kerr medium, which is modelled
by the anharmonic oscillator, is rather restricted \cite{Tanas1,Tanas2,Tanas3}.
But in general the problem,  even in the rather simple  case $|\alpha\rangle+|\beta\rangle+|\gamma\rangle$, is open under debate.
\section{Conclusion}
In summary, we have introduced generalized balanced N-mode  entangled coherent states
$|\Psi^{(d)}_{N}\rangle$ which can be recast as  multi-qudit pure states. To this aim we required  linearly independent of various coherent states appearing in superposition. For simplicity, we stick here  to pure and mixed states involving real parameters,  except for pure N-qubit states. It was shown that the state $|\Psi^{(d)}_{N}\rangle$ is separable if and only if all $\mu_{i}=0$.  To analyze the  entanglement of pure and mixed qubit like states, we used the concurrence measure, introduced by Wootetters,  while for general pure   N-mode cases we applied concurrence vector introduced by Akhtarshenas. For N-qubit pure states, the concurrence ${C^{(2)}_{m,N-m}}$ shows a significant enhancement by  increasing $m\leq\frac{N}{2}$ where the maximum is achieved for balanced partition, i.e. $m=\frac{N}{2}$. The same result holds for qutrit states which means that the concurrence $C^{(3)}_{m,N-m}$ has maximum value for  $p_i=0$ and  $\mu_{1,2}=\pm1$, whence the state reduces to GHZ state. We saw that, unlike in N-qubit cases, there are pure  N-qutrit states violating monogamy inequality.
Based on parity, displacement operators and beam splitters, we have proposed a protocol  for generating balanced N-mode entangled coherent states with even number of terms appearing in superposition.
\par
\textbf{Acknowledgments}\\
The authors also acknowledge the support from the University of Mohaghegh Ardabili.

\end{document}